\documentclass[preprint,12pt]{elsarticle}
\usepackage{pifont}
\usepackage{natbib}
\usepackage{geometry}
\usepackage{fleqn}
\usepackage{graphicx}
\usepackage{amssymb}
\journal{Computer Physics Communications}

\begin{document}

\begin{frontmatter}



\title{\textsc{TimeSeriesStreaming.vi}:\\LabVIEW program\\for reliable data streaming of large analog time series}
\author{Fabian Czerwinski}
\ead{czerwinski@nbi.dk}
\ead[url]{www.nbi.dk/\~{}czerwin}
\author{Lene B. Oddershede}
\address{Niels Bohr Institute, Blegdamsvej 17, 2100 Copenhagen, Denmark}
\begin{abstract}
With modern data acquisition devices that work fast and very precise,
scientists often face the task of dealing with huge amounts of data. These need to
be rapidly processed and stored onto a hard disk. We present a LabVIEW
program which reliably streams analog time series of MHz sampling. Its run time has virtually no limitation.
We explicitly show how to use the program to extract time series 
from two experiments: For a 
photodiode detection system that tracks the position
of an optically trapped particle and for a measurement of ionic current
through a glass capillary.
The program is easy to use and versatile as the input can be any type of analog signal. Also, the
data streaming software is simple, highly reliable, and can be easily
customized to include, e.g., real-time power spectral analysis
and Allan variance noise quantification.

%%%%%%%%%Block of information concerning program%%%%%%%%%%%%%%%%%%%%%%%%
\section*{Program summary}	
\noindent \textit{Program title:} \textsc{TimeSeriesStreaming.vi}\\
\textit{Catalogue identifier}: TimeSeriesStreaming\_v1\_0	\\
\textit{Program summary URL:} http://www.nbi.dk/\~{}czerwin/TimeSeriesStreaming.html\\
\textit{Licensing provisions:} Standard CPC licence, http://cpc.cs.qub.ac.uk/licence/licence.html\\
%\textit{No. of lines in distributed program, including test data, etc.:} tbd	\\
%\textit{No. of bytes in distributed program, including test data, etc.:} tbd	\\
\textit{Distribution format:} .zip / .vi	\\
\textit{Programming language:} LabVIEW	\\
\textit{Computer:} Any machine running LabVIEW 8.6 or higher\\
\textit{Operating system:} Windows XP and Windows 7\\
\textit{RAM:} 60 -- 360~Mbyte\\
\textit{Memory usage:} Code 40.5~Kbyte, front panel 50.7~Kbyte, block diagram 444.1~Kbyte; total VI file size 77.7~Kbyte\\
\textit{Classification:} Data acquisition program	\\
\textit{Nature of problem:} For numerous scientific and engineering applications, it is highly desirable 
to have an efficient, reliable, and flexible program to perform data streaming of time series sampled 
with high frequencies and
possibly for long time intervals. This type of data acquisition often produces very large amounts of data 
not easily streamed onto a computer hard disk using standard methods.\\
\textit{Solution:} This LabVIEW program is developed to directly stream any kind of time series onto a hard disk. Due to optimized timing and usage of computational resources, such as multicores and protocols 
for memory usage, this program provides extremely reliable 
data acquisition. 
In particular, the program is optimized to deal with large amounts of data, e.g., taken with high sampling
frequencies and over long time intervals. The program can be easily customized for time series analyses.\\
\textit{Restrictions:} Only tested in Windows-operating LabVIEW environments, must use TDMS format, 
acquisition cards must be LabVIEW compatible, driver DAQmx installed.\\
\textit{Running time:} as desirable: microseconds to hours\\

\end{abstract}
\begin{keyword}
data acquisition \sep data streaming \sep LabVIEW \sep TDMS \sep optical tweezers
\PACS 87.80.Cc \sep 87.80.Ek \sep 07.05.Hd \sep 07.90.+c
\end{keyword}
\end{frontmatter}

%%% main text
%%%%%%%%%%%%%%%%%%%%%%%%%% INTRO %%%%%%%%%%%%%%%%%%%%%%%%%%%
\section{Introduction}
Precision experiments where data is acquired with high temporal resolution pose a challenge with respect to
streaming and saving the data correctly onto a computer hard disk for further processing~\cite{NeumanRSI2004}. 
Within the nanoscience and biophysical communities, LabVIEW is often the program of choice for control of
data acquisition and streaming~\cite{OddershedePNAS2007, vanMamerenJPhysChemB2009, MahamdehOptExpress2009}. Here, we present a
highly reliable and efficient data streaming program in LabVIEW. The program is built into modular blocks with the goal
of making the design comprehensible and easily compatible for further customization. Also,
user-friendliness has been highly valued and we show how to use the program to stream time series data
from two typical nanoscale experiments: One
involving optical trapping assays~\cite{CzerwinskiOptExpress2009}, the other ionic current measurements through glass capillaries~\cite{SteinbockNanoLett2010}.

%%%%%%%%%%%%%%%%%%%%%%%% PROGRAM OVERVIEW %%%%%%%%%%%%%%%%%%%%
\section{Program Overview}

\subsection{Requirements}
A time series can originate from a wide range of physical signals, such as temperature, voltage, current etc. 
The time series, most often in the form of parallel voltage signals, enter the program through a number of 
channels of an acquisition card, building the interface between computers and experimental setups. Acquisition cards 
are available 
in a broad range for various tasks and quality requirements. We used National Instruments cards NI PCI-6251, NI PCIe-M6251, and NI PCI-M6040~\cite{SpecsNIAcqCards}. They are coupled 
into \textsc{TimeSeriesStreaming.vi} by DAQmx, a LabVIEW-internal driver. As precondition, the acquisition card must be compatible with LabVIEW. This holds either for those that can be installed by National Instrument's \emph{Measurement and Automation Explorer}, or for those supplied with a LabVIEW-compatible driver.

\subsection{Main Program}
The main program is designed in a modular fashion to offer independent as well as 
interconnected control of different sources of analog signals. Further, it contains support 
for data-streaming protocols. The programming architecture combines horizontal modules (acquisition, queuing, streaming) with vertical programming patterns (sequential structure, parallel while loops, multicore processing) in order to assure negligible error rates and optional customization.
\begin{figure*}[htbp]
\begin{center}
\includegraphics[width=\columnwidth]{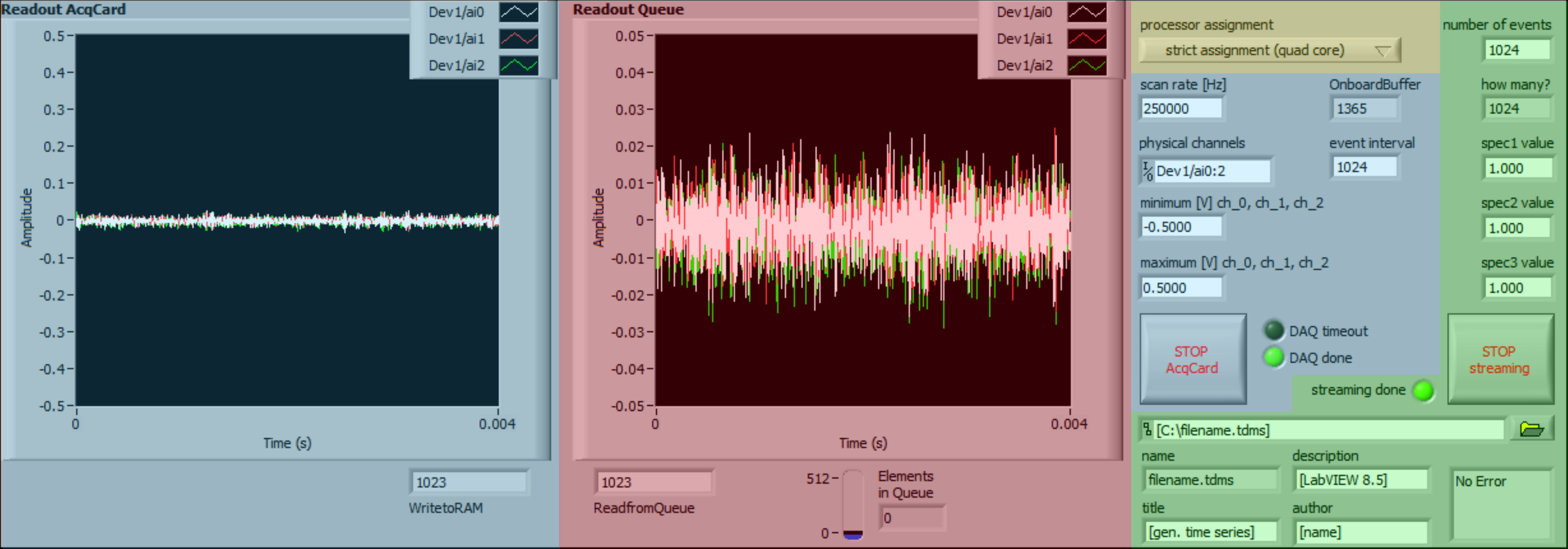}
\caption{Front panel of \textsc{TimeSeriesStreaming.vi}. In- and outputs of the four programming modules are highlighted by background color: \textbf{computer specifications} (yellow), \textbf{acquisition} (blue), \textbf{queuing} (red), and \textbf{streaming} (green). Details regarding the user defined settings are given in Section~\ref{sec:howto}. A high-resolution version of Figure~\ref{fig:FrontPanel} is available through http://www.nbi.dk/~czerwin/TimeSeriesStreaming.html.}
\label{fig:FrontPanel}
\end{center}
\end{figure*}

The different modules of the program are highlighted each by their background color in 
Figures~\ref{fig:FrontPanel} and \ref{fig:BlockDiagram}. The four modules deal with elements 
that concern \textbf{computer specifications} (yellow), \textbf{acquisition} (blue), \textbf{queuing} (red), and \textbf{streaming} (green).
Each of the modules functions independently from the others as it communicates through well-defined 
programming patterns. 

The usage of the program will be explained in Section~\ref{sec:howto}.

\paragraph{Programming Patterns}

\emph{Multicore processing} is the ability to distribute computational jobs over more than one core, i.e. one CPU. This 
feature has become available in recent versions of LabVIEW. In \textsc{TimeSeriesStreaming.vi} multicore processing is implemented 
by assigning each timed loop to a specific core. On the tested systems, the CPU load of an individual core never exceeded 20\%. Optimal multicoring was ensured by core assignments (highlighted yellow in Figure~\ref{fig:BlockDiagram}). It could also compensate for occasional interruptions by the Windows XP operating system.

Data acquisition must not be interrupted by waiting times during the streaming
process. LabVIEW is optimized for \emph{data flow control}. In \textsc{TimeSeriesStreaming.vi} this is achieved by 
transporting data packages between different loops exclusively through built-in queues (highlighted red
in Figure~\ref{fig:BlockDiagram}). The streaming loop is not executed when the queue is empty. 
This strategy has proven very powerful, as it allows both loops to run as quickly as possible without 
potential disturbance by waiting times.

\emph{Parallelizing} allows for parallel execution of computational jobs. Data acquisition is done in one while loop, 
data streaming in a parallel loop. Very reliable \emph{streaming} is achieved by the powerful data format TDMS 
(Technical Data Management Streaming, National Instruments). Using the primitive TDMS VIs allows for high performance 
streaming virtually with no limitation.

%%%%%%%%%%%%%%%%%%%%%%%%%%%%% HOW TO %%%%%%%%%%%%%%%%%%%%%%%%%%%%%%%%%%
\section{How to use the program} \label{sec:howto}
\begin{figure*}[htbp]
\begin{center}
\includegraphics[width=\columnwidth]{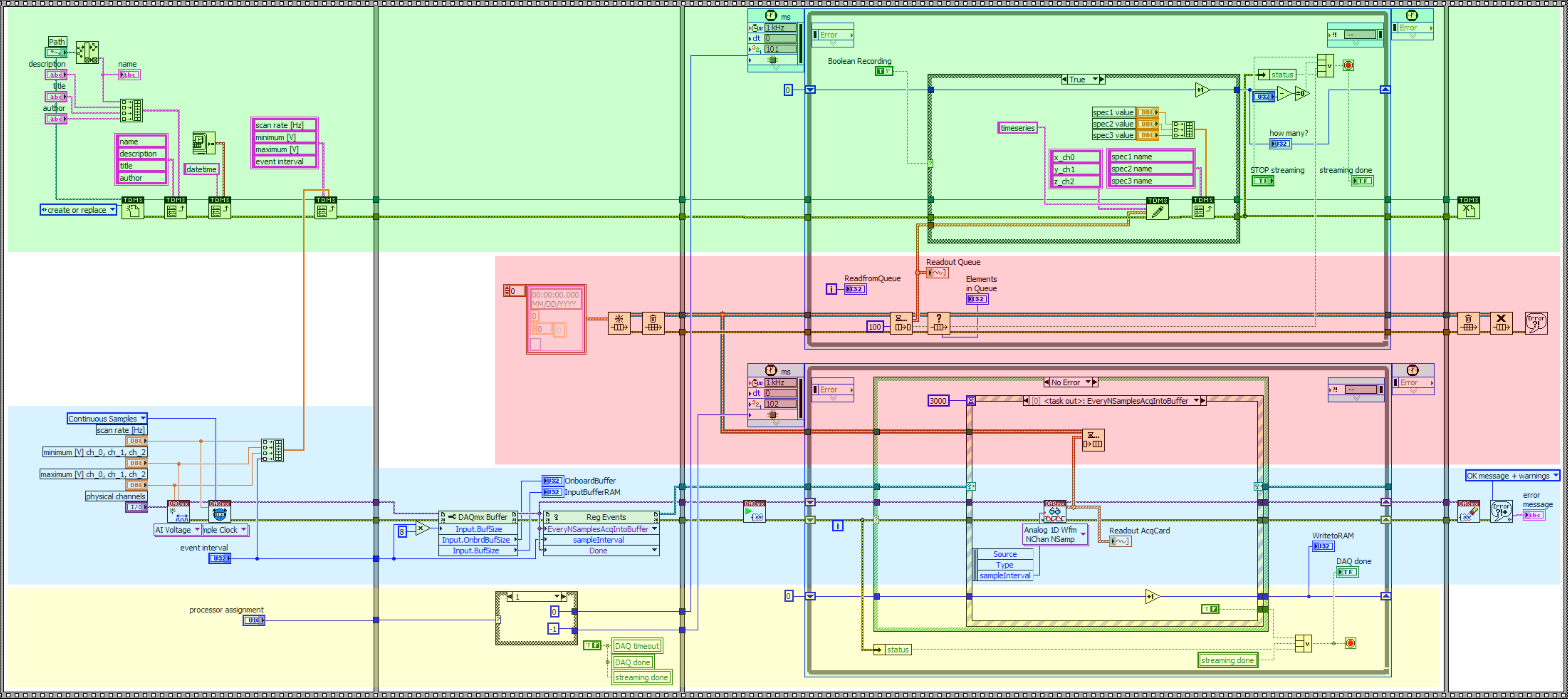}
\caption{Block diagram of \textsc{TimeSeriesStreaming.vi}. The four programming modules are highlighted by background color: \textbf{computer specifications} (yellow), \textbf{acquisition} (blue), \textbf{queuing} (red), and \textbf{streaming} (green). This diagram illustrates the modular architecture of the program where horizontal modules (acquisition, queuing, streaming) can be combined with vertical programming patterns (sequential structure, parallel while loops). A high-resolution version of Figure~\ref{fig:BlockDiagram} is available through http://www.nbi.dk/~czerwin/TimeSeriesStreaming.html.}
\label{fig:BlockDiagram}
\end{center}
\end{figure*}

A hands-on introduction to \textsc{TimeSeriesStreaming.vi} is given in this section. The perspective user is guided through 
the modules in the program's front panel (Figure~\ref{fig:FrontPanel}). An experienced user could adjust the programming 
architecture in the block diagram at will (Figure~\ref{fig:BlockDiagram}).

The only \textbf{computer specification} that must be set is the \textsf{processor assignment} (highlighted yellow). The user can 
choose to perform processor assignment on quad-, duo-, or single-cores; or simply choose \textsf{automatic} in which case the program 
will usually assign the highest ordered cores. However, if the user is aware, e.g., that the Windows operating system is utilizing 
certain cores, it might be beneficial to assign the cores manually.

The \textbf{acquisition} module (highlighted blue) controls mainly the settings regarding the acquisition card for a particular measurement. The desired \textsf{scan rate [Hz]} must be given in units of Hertz. The \textsf{physical channels} must be specified. \texttt{Dev1/ai0:2} 
denotes the signals input from Dev1 through the channels $0$, $1$, and $2$ (e.g., $x$, $y$, and $z$ coordinates of a recorded movement).
The internal limits of recordable voltage signal \textsf{minimum [V]} and \textsf{maximum [V]} are set in order to optimize the resolution of the recorded time series. It is advisable to have these settings as close as possible to the extrema of the input time series, though, without cutting any of the data points. 

Acquisition cards are equipped with an on-board buffer of a certain size indicated by \textsf{OnboardBuffer}. \textsf{event interval} sets the number of data points in each interval. If this number does not exceed the limit given by \textsf{OnboardBuffer}, the data are optimally passed onto the memory. Therefore, it is recommendable to try to keep \textsf{event interval} smaller than \textsf{OnboardBuffer}.

On left side of the front panel, a graph displays the output of the acquisition card in Volts for all specified channels. 
The graph \textsf{Readout AcqCard} shows the last package passed to the memory. The values of \textsf{minimum [V]} and \textsf{maximum [V]} 
shall be set as vertical axis limits. At any time, the acquisition can be aborted by hitting \textsf{STOP AcqCard}. This will halt the 
execution of looped functions of the DAQmx driver. The indicators \textsf{DAQ timeout} and \textsf{DAQ done} light up, if one of 
these two reasons terminates loop and therewith the program.

The graph \textsf{Readout Queue} (highlighted red) displays the last package of data passed through the \textbf{queue} to 
the streaming module. Utilizing a user-specified voltage interval here allows for an on-screen check, e.g., for whether data could 
be exposed to drift. The indicator \textsf{Elements in Queue} reveals the queue's filling level. 

\textbf{Streaming} and data storage are controlled by the settings highlighted in green. With \textsf{number of events} the user specifies the total number of event intervals to be acquired. Hence, the total number of data points will be:
$$
\textsf{total number of points}=\textsf{number of events} * \textsf{event interval}\textnormal{.}
$$
\textsf{how many?} counts the processed number of events. The entries \textsf{spec1 value}, \textsf{spec2 value}, and
\textsf{spec3 value} are numerical values. They will be stored in the resulting TDMS file in the header to the recorded data. It can regard values the user wishes to keep with the data (e.g., laser power, particle dimensions, ...).

The exact path for the TDMS file with the recorded data must be given. Therefore, replace \texttt{[C:\textbackslash filename.tdms]} by an 
appropriate entry. Note that an already existing file will be overwritten by default. The program automatically returns the file \textsf{name} and stores it in the header of the TDMS file. \textsf{title}, \textsf{description}, and \textsf{author} can be defined by the user 
and are also stored in the file header. 

If the total number of data points has been streamed, or if the \textsf{STOP streaming} button has been pressed, the program terminates and the indicator \textsf{streaming done} lights up. All data recorded up to that point is available in the user defined TDMS file. Also, TDMS files could be access already during streaming, e.g., from inside another VI. 

Potential error messages are shown in the bottom right box.

\section{Data Format and Benchmarks}
\paragraph{Data Format}
The data recorded with \textsc{TimeSeriesStreaming.vi} is stored in a TDMS file. TDMS is an open-source file format developed by National Instruments. It is a binary format optimized for data streaming, thus, it 
handles dynamically increasing files correctly. Also, one can access the file already during acquisition. 
There are three ways to access TDMS: 
Directly in LabVIEW using the primitive TDMS VIs, by third-party plug-ins, or by the program \textit{Diadem} (National Instruments). 
For the latter method, it is essential to use \textit{Diadem} version 11.1 or higher. Detailed information on how to use Diadem or how 
to export TDMS files into third-party products can be found under http://www.ni.com/tdms. Third-party plug-ins exist, e.g., for \textit{Matlab}, \textit{OpenOffice}, and \textit{Excel}. \textit{Origin} can open TDMS files directly. 

\paragraph{Benchmarks}
We tested the performance and stability of \textsc{TimeSeriesStreaming.vi} on three different computers designated for data
acquisition. Their specifications are made available through the program summary URL~\cite{online}. Since LabVIEW only provides acceptable performance on machines operating Windows, we were limited by the choice of the operating system. All tested CPU architectures 
(quad-, duo-, and single-core) proved to yield very good acquisition. Though, the highest level 
of customizability applied to multicore processors. We tested the acquisition cards NI PCI-M6040, NI PCI-6521, and NI PCIe-6521.
\begin{table*}[htdp] 
\begin{center}
\begin{tabular}{rrrrrr}
\multicolumn{1}{c}{scan rate}&\multicolumn{1}{c}{number of}&\multicolumn{1}{c}{buffer per}&\multicolumn{1}{c}{max filling}&\multicolumn{1}{c}{TDMS file size}&\multicolumn{1}{c}{error rate}\\
\multicolumn{1}{c}{(kHz)}&\multicolumn{1}{c}{channels}&\multicolumn{1}{c}{channel}&\multicolumn{1}{c}{of queue	}&\multicolumn{1}{c}{(MB/s)}&\\
\hline
1,250	&1			&4096				&0			&9.6				&$< .000$\\
500		&2			&2048				&0			&7.7				&$< .000$\\
250		&4			&1024				&0			&15.3			&$.004$\\
100		&3			&1366				&0			&2.3				&$.001$\\
100		&1			&4096				&2			&2.3				&$.001$\\
22		&6			&683				&5			&6.0				&$.003$\\
22		&3			&1366				&3			&3.0				&$.001$\\
22		&1			&4096				&3			&1.0				&$.001$\\
10		&1			&4096				&8			&0.5				&$< .001$\\
1		&1			&4096				&3			&0.05			&$< .000$\\
\end{tabular}
\caption{Benchmark of \textsc{TimeSeriesStreaming.vi} as performed on a NI PCI-6251 connected to a Windows XP computer running LabVIEW 8.6. Each individual run took 4096~s. Scan rate, number of channels, and buffer per channel were set. Maximum filling, file size, and error rate were determined.}
\end{center}
\label{default}
\end{table*}%
Table 1 represents the benchmarks for the NI PCI-6251 card
with thermal noise as input. Scan rate, number of channels, and buffer per channel were set. Then the program ran for 4096~s and maximum filling and file size were determined. 
An error was counted when two consecutive 16-bit digits were exactly identical. The error rate was
defined as number of errors divided by the number of data points.

In addition, we simulated basically all voltage acquisition devices with the \textit{Measurement and Automation Explorer} and performed benchmarks with \textsc{TimeSeriesStreaming.vi} using the input from the simulated device.

%%%%%%%%%%%%%%%%%%%% EXAMPLES &&&&&&&&&&&&&&&&&&&&&&&&&&&&&&&&
\section{Examples of Program Applications}
Two brief examples of how to use \textsc{TimeSeriesStreaming.vi} are described here. In addition, the program has already been used 
to reliably stream large time series from experiments involving optical trapping of micron-sized polystyrene 
spheres~\cite{CzerwinskiOptExpress2009}, gold nanorods~\cite{CzerwinskiSPIEProc2009}, and quantum 
dots~\cite{JauffredSPIEProc2010}. Furthermore, we implemented an improved calibration protocol for optical tweezers that made use of the main programming features introduced here~\cite{AnderssonSUBM2010}.
\begin{figure}[htbp]
\begin{center}
\includegraphics[width=0.7\columnwidth]{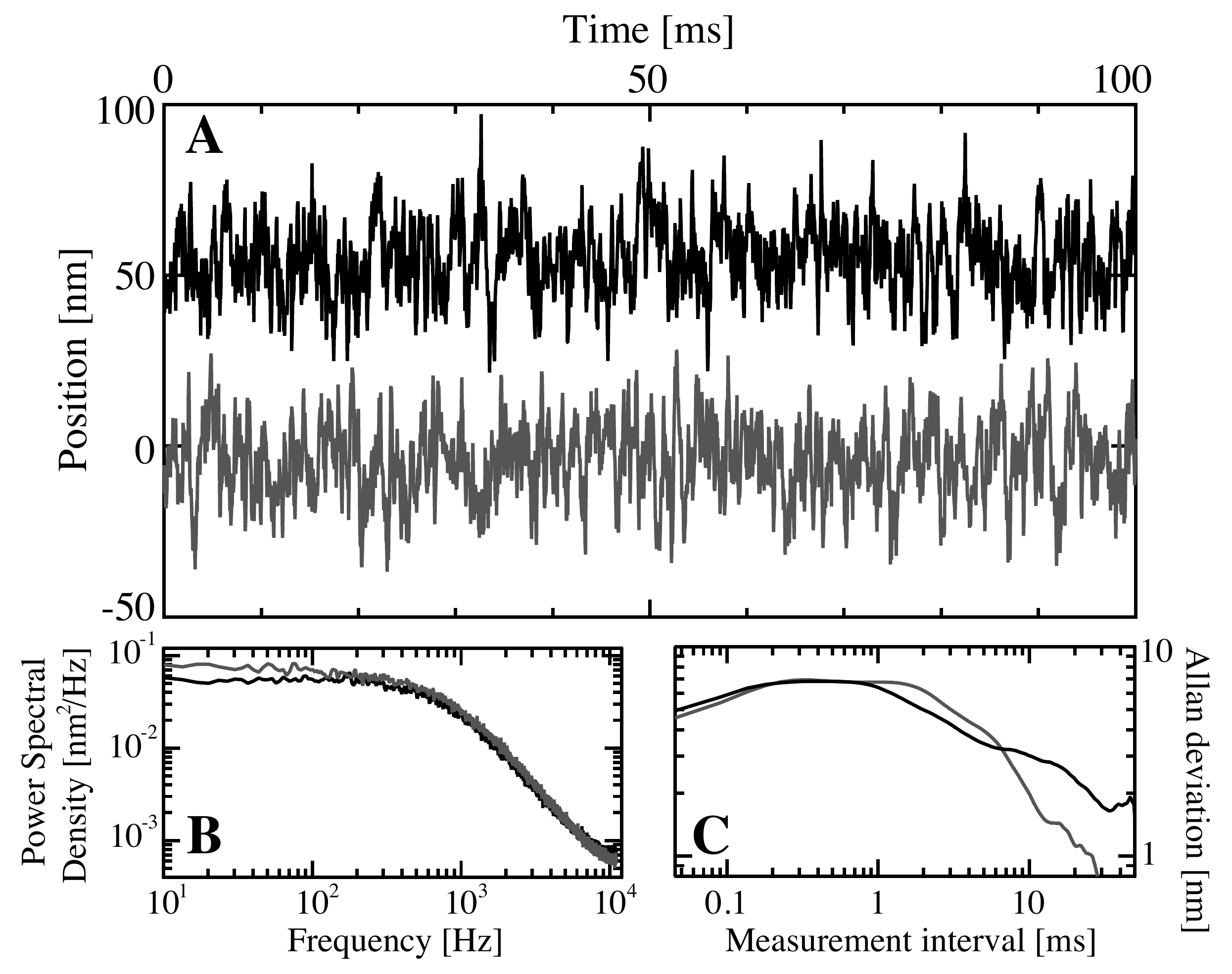}
\caption{Example of data acquisition and time series analyses applied to an optical trapping assay. \textbf{A} Positional time series of an 800~nm polystyrene sphere which is harmonically trapped. Its lateral positions $x$ (grey) and $y$ (black, offset 50~nm) are recorded by a photodiode detection system. The power spectral density (\textbf{B}) and the Allan deviation (\textbf{C}) of the lateral positions visited.}
\label{fig:motivation}
\end{center}
\end{figure}

\subsection{Position Recording in Optical Tweezers}
The development of \textsc{TimeSeriesStreaming.vi} was prompted by a need to stream large amounts of data from optical trapping assay to a hard disk. The goal was to analyze the noise by means of accuracy measurements~\cite{CzerwinskiOptExpress2009}. Therefore, positions of a trapped microsphere were recorded at sampling frequencies of up to 100~kHz in the order of hours. A short time series is plotted in Figure~\ref{fig:motivation}A.

Experimental details are provided in Reference~\cite{CzerwinskiOptExpress2009}. The positions were sampled using a photodiode 
detection system yielding an output in Volts, which were reformulated in terms of metric distances by a calibration factor. 
Figure~\ref{fig:motivation}B shows the positional power spectrum, which, when properly analyzed, gives information about
the calibration factor as well as the strength of the optical trap~\cite{BergSorensenRSI2004}. 
A different type of time series analysis, Allan variance analysis, is excellent for quantifying noise in optical trapping 
assays~\cite{CzerwinskiOptExpress2009, AnderssonSUBM2010}, in particular in the
low frequency regime, which is not possible through normal variance or power spectral analysis. 
Figure~\ref{fig:motivation}C shows the Allan 
deviation of the same trace quantifying the exact accuracy for various measurement intervals. 
For this type of analysis it is crucial to have long overall measurement time series and reliable
streaming of the data onto the hard disk, a requirement met by \textsc{TimeSeriesStreaming.vi}.  
The modular fashion of the program enables 
straight forward implementation of similar types of calibration or noise quantification routines.

\subsection{Ionic Current through Glass Capillary}
\begin{figure}[htbp]
\begin{center}
\includegraphics[width=0.7\columnwidth]{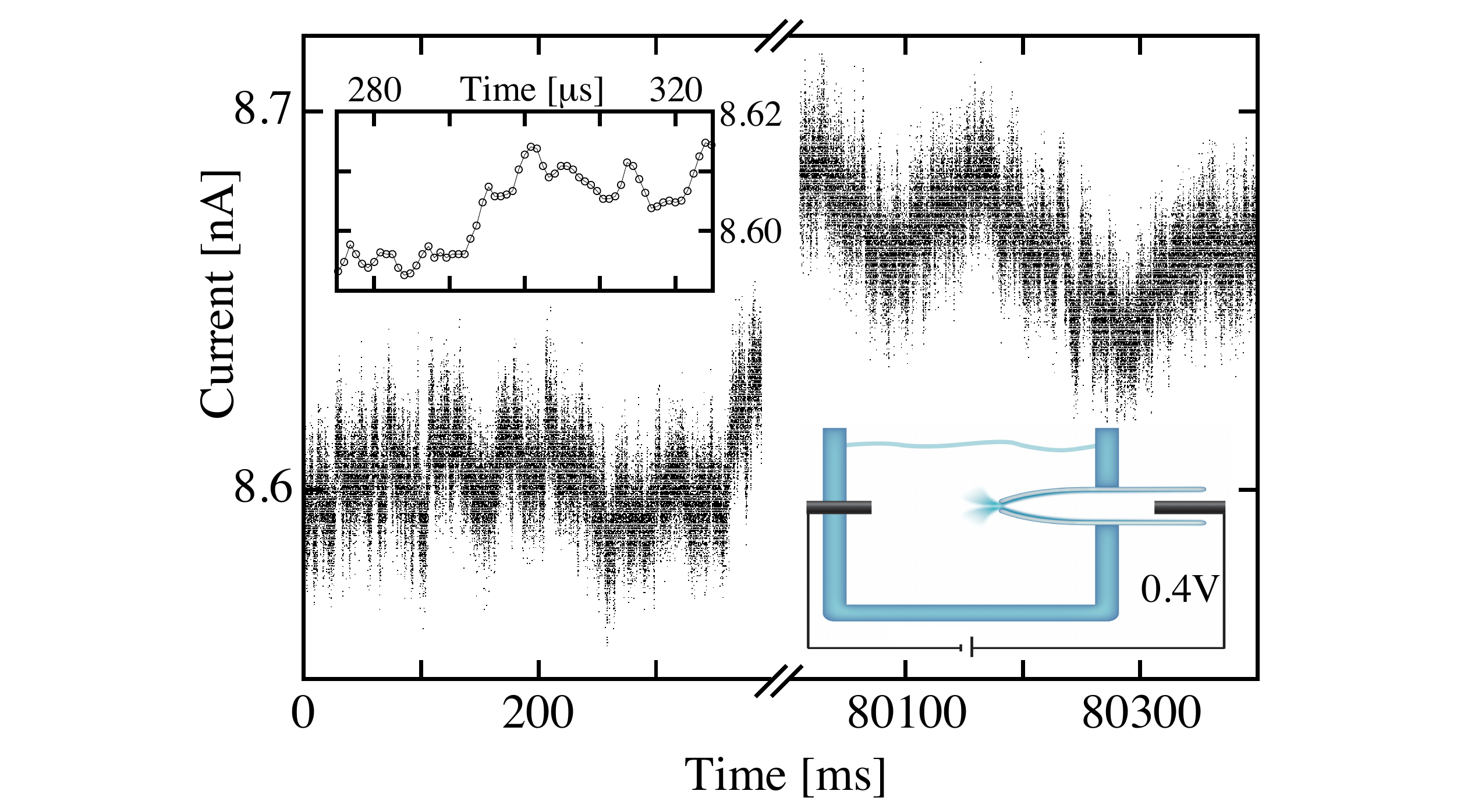}
\caption{Time series from measurements of monovalent-ionic current through a glass capillary (diameter 60~nm), a sketch
of the experiment is shown in the lower right. The inset in the upper left is a zoom-in on the time axis,
showing that the acquisition card's limit of 1.25~MHz sampling can be achieved.}
\label{fig:CurrentTrace}
\end{center}
\end{figure}

The translocation of molecules through solid-state nanopores has drawn a lot of attention in recent years due to the enormous 
potential they hold for parallel screening of biomolecular solutions~\cite{DekkerNatureNanotech2007}. Also, 
glass capillaries with a diameter of 60~nm could be used to detect DNA folding~\cite{SteinbockNanoLett2010}. Here, we 
used \textsc{TimeSeriesStreaming.vi} to stream the the ionic current measured onto the computer hard disk.
A sketch of the experiment is shown in the lower right of Figure~\ref{fig:CurrentTrace}.
The measured current values are plotted, and in the upper left there is a 
zoom-in to illustrate that data acquisition was done at a very high rate (1.25~MHz). In this
experiment, the error rate was zero.
Hence, time series can be analyzed for events happening within sub-milliseconds, 
the timescale relevant for protein translocation through a nanopore~\cite{DekkerNanoLett2010}, or for events
happening on the order of minutes, a typical timescale for drift, as also visible on Figure~\ref{fig:CurrentTrace}.

\section{Summary}
We developed a LabVIEW program that reliably streams large amounts of data correctly onto a computer hard disk. The program 
was checked on several individual platforms and showed to perform with a very small error rate.
The program is made in a modular fashion with the aim of making it user-friendly and easily customizable. As an example, we showed how to acquire the positions of an optically trapped sphere and how this time series data can be further analyzed. We also demonstrated how the program streamed time series of ionic current measurements. 
As the program easily handles a broad range of analog inputs, there is a wide range of applications, particularly in biophysical 
nano-scale experiments as pointed out in the two examples. 
The source code is freely available at through the CPC Program Library and under the standard CPC license agreement~\cite{license}.

\section*{Acknowledgements}
We thank Lorenz Steinbock for data acquisition and help on the ionic current measurements, and Oliver Otto as well as Jesper Tholstrup for comments on the manuscript. Also, we acknowledge financial support through the KU excellence program.

\bibliographystyle{elsarticle-num}

\end{document}